# Long-lived quantum coherence in photosynthetic complexes at physiological temperature

Gitt Panitchayangkoon[1], Dugan Hayes[1], Kelly A. Fransted[1], Justin R. Caram[1], Elad Harel[1], Jianzhong Wen[2], Robert E. Blankenship[2] & Gregory S. Engel[1]

[1]*Department of Chemistry and The James Franck Institute, The University of Chicago, Chicago, Illinois 60637 USA*

[2]Departments of Biology and Chemistry, Washington University, St. Louis, Missouri 63130, USA

Corresponding author:

Prof. Gregory S. Engel

Gordon Center for Integrative Science

929 E 57th St, GCIS E119

Chicago, IL 60637

phone: (773) 834-0818

e-mail: gsengel@uchicago.edu

**Abstract**

Photosynthetic antenna complexes capture and concentrate solar radiation by transferring the excitation to the reaction center which stores energy from the photon in chemical bonds. This process occurs with near-perfect quantum efficiency. Recent experiments at cryogenic temperatures have revealed that coherent energy transfer – a wavelike transfer mechanism – occurs in many photosynthetic pigment-protein complexes (1-4). Using the Fenna-Matthews-Olson antenna complex (FMO) as a model system, theoretical studies incorporating both incoherent and coherent transfer as well as thermal dephasing predict that environmentally assisted quantum transfer efficiency peaks near physiological temperature; these studies further show that this process is equivalent to a quantum random walk algorithm (5-8). This theory requires long-lived quantum coherence at room temperature, which never has been observed in FMO. Here we present the first evidence that quantum coherence survives in FMO at physiological temperature for at least 300 fs, long enough to perform a rudimentary quantum computational operation. This data proves that the wave-like energy transfer process discovered at 77 K is directly relevant to biological function. Microscopically, we attribute this long coherence lifetime to correlated motions within the protein matrix encapsulating the chromophores, and we find that the degree of protection afforded by the protein appears constant between 77 K and 277 K. The protein shapes the energy landscape and mediates an efficient energy transfer despite thermal fluctuations. The persistence of quantum coherence in a dynamic, disordered system under these conditions suggests a new biomimetic strategy for designing dedicated quantum computational devices that can operate at high temperature.

**Introduction**

The FMO pigment-protein complex from *Chlorobium tepidum* serves as a model system for photosynthetic energy transfer processes (1, 9-12). This complex conducts energy from the larger light-harvesting chlorosome to the reaction center in green sulfur bacteria (13, 14). Each non-interacting FMO monomer contains seven coupled bacteriochlorophyll-*a* chromophores arranged asymmetrically, yielding seven non-degenerate, delocalized molecular excited states called excitons (10, 15). The small number of distinct states makes this particular complex attractive for theoretical studies of transport dynamics. As shown by Ishizaki et al., the arrangement of the chromophores in FMO results in a downhill, rugged energetic landscape with two distinct routes through which an excitation can travel to reach the lowest energy state (12). While classical trajectories can navigate such funnel-like landscapes, the wavelike motion through the complex improves efficiency by avoiding kinetic traps. In higher plants, this wavelike mechanism likely becomes more important because the landscape is more rugged without a downhill arrangement (16).

Recent investigations of photosynthetic systems at 77 K have found evidence of coherent energy transfer in many antenna complexes and even in the reaction center of purple bacteria (1-3). This wavelike energy transfer mechanism, however, can only contribute to the near perfect quantum efficiency of photosynthesis if coherence survives in these systems during energy transfer at physiological temperatures. As temperature increases, thermally excited vibrational modes of the protein bath drive larger energetic fluctuations, thereby accelerating decoherence (13, 17). Although this dephasing seems unfavorable, Rebentrost et al. and Plenio and Huelga have independently shown that the delicate interplay between quantum coherence and dephasing can create fast and unidirectional transfer pathways within these complexes, resulting in highly efficient electronic energy transfer (5-8, 18). This scheme exploits quantum coherence to overcome an energy barrier, but subsequent dephasing processes trap the excitation at the target site. Optimal transport therefore requires both dephasing and coherent energy transfer.

The initial excitation or transfer event necessarily creates quantum coherence because both the dipole and site operators do not commute with the system Hamiltonian. For a system of two excitons described by $\Psi(t) = c_1\phi_1 + c_2\phi_2$, the time evolution of the density matrix is given by

$$|\Psi(t)\rangle\langle\Psi(t)| = |c_1|^2|\phi_1\rangle\langle\phi_1| + |c_2|^2|\phi_2\rangle\langle\phi_2| + c_1 c_2^* e^{-i(E_1-E_2)t/\hbar}|\phi_1\rangle\langle\phi_2| \\ + c_1^* c_2 e^{i(E_1-E_2)t/\hbar}|\phi_2\rangle\langle\phi_1|. \quad (1)$$

The first two terms represent populations in the excitonic basis, while the latter two describe coherences. The phase factors in the coherence terms are responsible for quantum beating, which appears as a periodic modulation of peak amplitude and population in the site basis. The frequency of this beating corresponds to the energy difference between the two excitons giving rise to that particular quantum coherence. Traditionally this phenomenon is ignored in transport dynamics because fast electronic dephasing generally destroys quantum coherence before it can impact the transport process. For example, at cryogenic temperature, coherences between ground and excited states in FMO dephase in approximately 70 fs. In contrast, coherences among excited states have been shown to persist beyond 660 fs – long enough to improve transport efficiency (1). Such a coherence can only persist if the electronic spectral motion among chromophores is strongly correlated, as demonstrated in a conjugated polymer system by Collini and Scholes (19). Biologically, these correlations arise because the protein environment forces transition energies of chromophores to fluctuate together due to spatial uniformity of the dielectric bath. The resultant long-lived quantum coherences evolve in time, creating periodic oscillations in both spectral signals and wavepacket position. This quantum beat provides the signature of quantum coherence.

Experimentally, we probe quantum coherences in FMO using two-dimensional Fourier transform electronic spectroscopy to directly observe electronic couplings and quantum coherences as a function of time (20-23). The experimental method and theory have been described in detail elsewhere (22). In short, three laser pulses interact in the weak field limit with the sample to produce a third-order polarization (24). The first pulse creates a superposition of ground and excited states, and the phase of this coherence evolves for a time $\tau$ (coherence time) before a second pulse creates either a population or a superposition of excited states for a time $T$ (waiting time). During this time, both population and coherence transfer occur, and coherences evolve phase. A third pulse then generates a second ground-excited state coherence. Finally, the signal pulse appears after a time $t$ (rephasing time) and is heterodyne-detected with a local oscillator pulse in a unique phase-matched direction. A Fourier transform of the signal along the coherence and rephasing time dimensions at a fixed waiting time gives a 2D spectrum that correlates the "input" coherence frequency to an "output" rephasing frequency. Cross-peaks (off-diagonal peaks) appear in this spectrum due to coupling between states, providing evidence of energy transfer.

Two types of response pathways – rephasing and non-rephasing – contribute to the overall signal in 2D spectroscopy. We probe these two pathways individually by switching the order of the first two pulses, which controls the direction in which the phase of the coherence evolves during the coherence time. In the rephasing (non-rephasing) pathway, phase evolution proceeds in the opposite (same) direction during the coherence and rephasing times, resulting in a photon echo (free induction decay) signal. When analyzed independently, the rephasing and non-rephasing signals provide complementary information on coherences because quantum beats during the waiting time appear in different positions in the spectra (25, 26). For rephasing pathways, beating appears in the cross-peaks, while for non-rephasing pathways it appears in the peaks along the diagonal. The sinusoidal beating pattern observed in a cross-peak consists of a single frequency corresponding to the transition energy gap between the two excited states that give rise to this peak. The quantum beating observed in a diagonal peak, however, is the sum of several different frequencies arising from coherences with all other states, which complicates attempts to measure individual quantum coherence lifetimes. We therefore choose to focus on the off-diagonal peaks to enable more accurate measurement of the lifetime of an individual quantum coherence.

**Results**

Representative 2D spectra in Fig. 1a-d show the real part of the third-order nonlinear response taken at four different temperatures (77 K, 125 K, 150 K, and 277 K) at a waiting time $T$ = 400 fs. The lowest energy peak is well resolved at low temperatures, but more vibrational modes of the protein are occupied at 277 K, resulting in faster dephasing between the ground and excited states. According to the time-bandwidth product, spectral resolution is inversely related to signal lifetime, so the short duration of the signal along $\tau$ and $t$ results in loss of spectral resolution in the respective frequency dimensions. Furthermore, the spectrum acquired from the flowing sample also suffers from rearrangement of the hydration shell around the protein within the first 70 fs of waiting time, causing a rapid growth in the anti-diagonal linewidths. In contrast, the low temperature spectra maintain resolution and narrow peak shapes even after 1 ps because the translational motions of the solvent are frozen.

Fig. 1-e shows the absolute value of the amplitude of the highlighted cross-peak as a function of waiting time for each temperature. This feature was chosen because it is well separated from the congested diagonal peaks and the amplitude therefore beats with only a single frequency. The spectral coordinates of the cross-peak were chosen according to the energies of excitons 1 & 3 taken from the Hamiltonian calculated by Adolphs and Renger (27). According to the projection-slice theorem, we can separate the third-order response into real (absorptive) and imaginary (dispersive) portions by fitting the phase of the signal to the pump-probe data taken at the same waiting time as described by Brixner et al (22). Separating the data into real and imaginary parts will improve spectral resolution by eliminating wide dispersive features. However, scatter from the frozen samples creates an interferometric signal that cannot be separated from the pump-probe signal; this interference introduces errors in the real portion of the phased 2D datasets that could be mistaken for quantum beating. Therefore, we present the peak amplitude in absolute value to eliminate possible phase errors. Examining the cross-peak amplitudes at all four temperatures, we see clear evidence of quantum beating. The beating signals demonstrate excellent agreement in both the phase and frequency across all temperatures, indicating that the same phenomenon discovered at 77 K extends to at least 277 K. The dephasing rate does, however show strong temperature dependence with coherence extending only to about 300 fs at 277 K, approximately four times faster than the dephasing at 77 K.

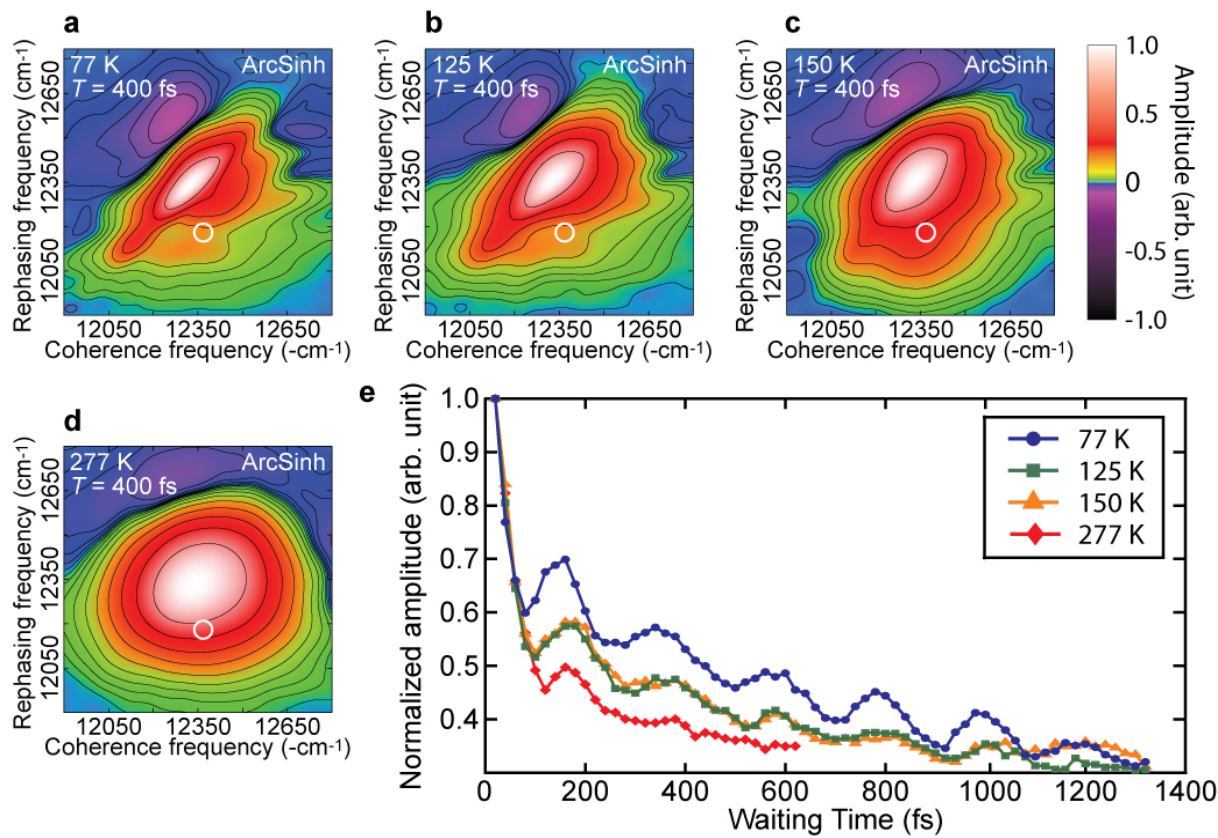

*Figure 1 | **Temperature dependence data**. Representative two-dimensional electronic spectra of FMO are shown at the waiting time T = 400 fs at 77 K (a), 125 K (b), 150 K (c), and 277 K (d). The data are shown with an arcsinh color scale to highlight small features in both negative and positive portions of the real third-order nonlinear response. Peaks broaden at higher temperature due to faster dephasing between ground and excited states, preventing resolution of the lowest excited state. The quantum beat signals are extracted at the spectral position (white circle) corresponding to the location of the 1-3 cross-peak predicted by a theoretical study(27). The beating signals (e) demonstrate agreement in phase and beating frequency among all 4 temperatures while showing shorter quantum beat lifetimes at higher temperatures.*

**Discussion**

Extracting the beating signal from the 277 K spectrum presents an additional challenge because of the broad lineshapes. Even using only the absorptive portion of the response to improve resolution, we cannot resolve individual cross-peaks in the spectra as illustrated in Fig. 2. With such broad lineshapes, beating from neighboring peaks confounds attempts to isolate a single quantum beating signal. We therefore isolate rephasing and non-rephasing signals separately both on and off the main diagonal to demonstrate that the beating occurs at the cross-peak only in the rephasing portion of the signal as predicted by theory (26). The Feynman diagrams in Fig. 2 show the Liouville-space pathways giving rise to the cross-peak signal in both rephasing and non-rephasing data indicated by the red and green arrows, respectively. While both rephasing (red) and non-rephasing (green) pathways produce signal at the same spectral position, these two pathways yield significantly different information during the waiting time *T*. In the rephasing pathway, we observe quantum beating on the cross-peak due to the phase factor $e^{i(E_3-E_1)t/\hbar}$ in the Green's function. The expected frequency of the beating is $\omega_{31} = \frac{E_3 - E_1}{\hbar}$. In contrast, the Green's function of the cross-peak during waiting time in the non-rephasing pathway equals 1, which indicates that it does not beat.

To verify that we have extracted actual quantum beating, we first contrast the rephasing signal on the main diagonal to the off-diagonal signal. The amplitude of the rephasing diagonal peak (solid green line) shows a smooth decay over the waiting time *T*, while the rephasing cross-peak amplitude (solid red line) shows multiple periods of quantum beating as it decays. Second, we compare the beating in the rephasing signal to the non-rephasing signal at the cross-peak (dashed red line). The non-rephasing signal in the region of the cross-peak demonstrates similar population dynamics to the rephasing signal but shows no quantum beating. These data confirm that the beating observed at room temperature arises from an electronic coherence, despite the loss of spectral resolution.

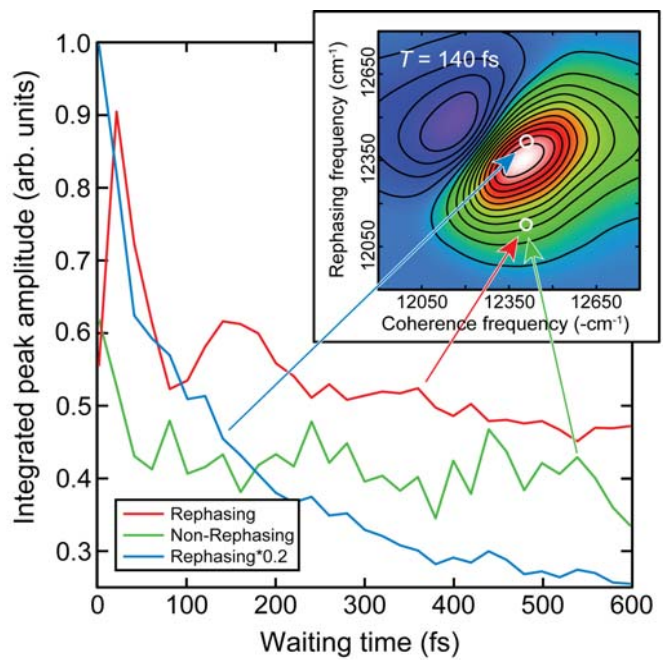

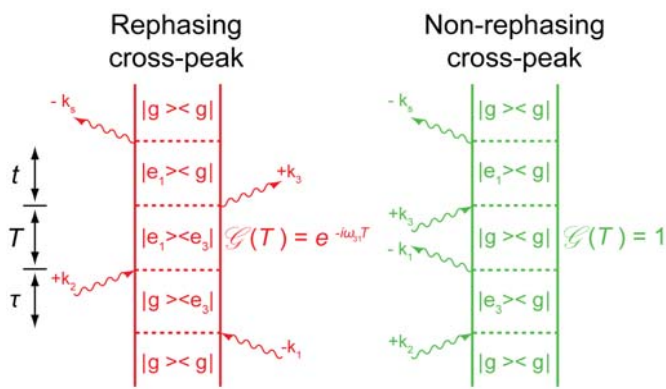

*Figure 2 | **Room temperature quantum beating.** Integrated peak amplitudes taken from the absorptive part of the 2D signal at 277 K are plotted as a function of waiting time. The rephasing signal at the indicated cross-peak (red line) shows multiple periods of quantum beating as it decays, while the non-rephasing signal (green line) shows no beating. As a further comparison, the rephasing signal at the indicated diagonal position (blue line) shows only smooth population decay. The magnitude of the diagonal signal was scaled by a factor of 0.2 to facilitate comparison with the off-diagonal signal. The inset shows the 2D spectrum of the absorptive rephasing signal taken at T = 140 fs. Below the spectra, Feynman diagrams representing the relevant Liouville pathways illustrate why beating arises only in the rephasing cross-peak signal.*

After elimination of population transfer dynamics from the beating signal by subtraction of a multi-exponential fit, we extract the beating signals from the coherence between exciton 1 & 3 shown in Fig. 3. Each signal is fit with the product of a sine function and an exponential decay from which the dephasing rate can be estimated. All regressions start from 80 fs to avoid pulse overlap effects and contributions from solvent reorganization at room temperature. The beating frequency found for each fit was approximately 160 cm$^{-1}$; in comparison, we would predict the frequency to be 198 cm$^{-1}$ corresponding to the energy difference between excited stated 1 and 3 given in the Hamiltonian referenced earlier. The dephasing rate taken from the exponential part of the fitting function is plotted as a function of temperature in the inset of Fig. 3. In comparison, Aspuru-Guzik and coworkers use an Ohmic spectral density with cut-off to model the FMO protein bath and predict the dephasing rate as a function of temperature to be $\gamma_\Phi(T) = 2\pi \frac{k_B T E_R}{\hbar^2 \omega_c}$ where $E_R$ is a reorganization energy (35 cm$^{-1}$) and $\omega_c$ is the cut-off frequency (150 cm$^{-1}$) (7). The observed linear relationship demonstrates that an Ohmic spectral density is a reasonable approximation for modeling this complex(28). Furthermore, the slope from the linear regression fit is 0.52 ± 0.07 cm$^{-1}$/K, agreeing with the slope obtained from the model.

Furthermore, Aspuru-Guzik and coworkers found that at room temperature the dephasing frequency is on the order of the excitonic energy gaps and couplings. This temperature corresponds to optimal transfer efficiency because dephasing traps an excitation at an energetic minimum, but does not collapse a coherence before it has a chance to complete at least a period of quantum beating and overcome the initial energy barrier. Intuitively, for a complete sampling of all possible pathways, all quantum coherences should have a chance to complete at least one oscillatory cycle. While coherence prevents loss to local minima, fast dephasing prevents loss to exciton recombination and non-radiative decay.

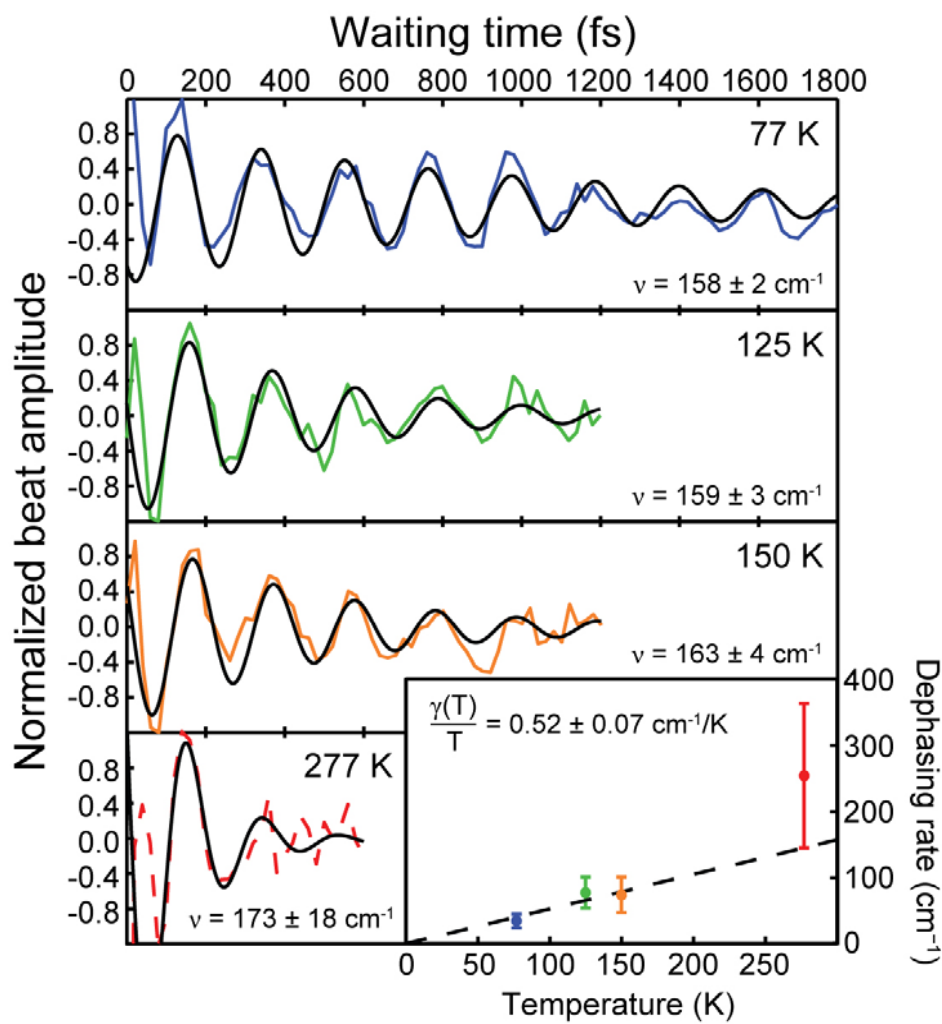

*Figure 3 | Temperature dependence of coherence dephasing. Integrated cross-peak amplitudes are taken from the absolute value of the combined 2D signal (rephasing + non-rephasing) after removal of exponential population decay at 77 K, 125 K, and 150 K (colored solid lines). The amplitude at 277 K is taken from the absorptive portion of the rephasing signal (dashed red line). The beating signals are normalized to their respective maxima and fit to the product of a sine function and an exponential decay (solid black lines). The beating frequency is given for each temperature. The dephasing rate taken from the exponential part of the fit is plotted as a function of temperature along with standard errors in the inset. The statistically weighted linear fit of these points (dashed black line) has a slope of 0.52 ± 0.07 cm-1/K.*

Thus, optimal efficiency is attained when dephasing completely destroys coherences shortly after a full period of quantum beating. Our experimental results show that not only does quantum coherence persist at physiological temperature long enough to impact overall transport dynamics, but it lasts for nearly two full cycles corresponding to the optimal transport efficiency. Therefore, this data supports the hypothesis that energy transfer in FMO has been tuned by natural selection to promote photosynthesis under natural conditions.

**Conclusion**

These data prove the same quantum beating signals observed at 77 K persist to physiological temperature and show agreement in both phase and frequency, indicating that the experiment is following the same quantum coherence at all temperatures. We observe a 130 fs e-folding lifetime for this excited state coherence at 277 K and observe quantum coherence lasting beyond 300 fs, showing that evolution has had the opportunity to exploit the theorized environmentally assisted quantum transport (EnAQT) mechanism for biological function. Beating may persist even longer, but we cannot separate actual dephasing from deleterious interference from other spectral features. Thus, this measurement represents a lower bound. Microscopically, this beating survives because the energies of the excited states involved fluctuate such that the energy gap remains largely constant. We hypothesize the fluctuations of the protein dielectric provide the long range spatial correlations on the appropriate timescale to protect quantum coherence and enable coherence transfer. As suggested by Cheng et al., this wavelike transfer mechanism results in oscillatory population dynamics, allowing particular sites to have momentary populations higher than their respective equilibrium populations (25). When coupled to an energy trap such as the reaction center, this process can greatly increase the quantum yield of a system relative to the classical limit. The long-lived quantum coherence in this complex also provides a model for non-unitary quantum algorithms at room temperature, which while not scalable may provide improvements beyond the classical limit for some operations.

**Materials and Methods**

**Sample preparation.** FMO from *Chlorobium tepidum* (29) in a buffer of 800mM tris/HCl (pH 8.0) was mixed 35:65 v/v in glycerol with 0.1% laurydimethylamine oxide detergent and loaded into a 200 μm fused quartz cell (Starna). The optical density at 809 nm was 0.32. A cryostat (Oxford Instruments) was used to cool the sample to 77K, 125 K and 150 K. The room temperature sample was prepared by mixing the sample 35:65 v/v in 800 mM tris/HCl buffer (pH 8.0) and pumped (Masterflex, Cole-Palmer) through a 200 μm quartz flow cell (Starna). The sample reservoir was kept in a water bath at 4 °C to cool the sample and prevent thermal degradation. The optical density of the room temperature sample at 809 nm was 0.35.

**Data Acquisition.** A self mode-locking Ti:Sapphire oscillator (Coherent, Micra) was used to seed a regenerative amplifier (Coherent, Legend Elite), which produced a 5.0 kHz pulse train of 38 fs pulses centered at 806 nm with a spectral bandwidth of 35 nm. The 10 Hz stability of the laser power during data acquisition ranged from 0.08% to 0.19%. Two pairs of phase-locked beams were generated by a diffractive optic (Holoeye). All beams were incident on identical optics except for one-degree fused silica wedges (Almaz Optics) on delay stages (Aerotech) which determined the time delays. Each delay was calibrated using spectral interferometry as described earlier (22, 30).

Neutral density filters with total optical density 3.1 at 809 nm attenuated the local oscillator beam. The total power incident on the sample was 4.8 nJ (1.6 nJ/pulse), which was focused to a spot size less than 70 μm. A 0.3m spectrometer (Andor Shamrock) frequency-resolved the emitted signal and local oscillator beam, which were captured on a 1600 x 5 pixel region of a back-illuminated, TE-cooled CCD (Andor Newton). Scatter subtraction, Fourier windowing, and transformation to frequency-frequency space were performed as described previously (22).

2D data were collected at waiting times (*T*) in 20 fs increments for all temperatures. At each waiting time, the coherence time was scanned from -500 to 500 fs in steps of 4 fs. A 2D spectrum at $T$ = 0 fs was taken every 200 fs to monitor sample integrity. No degradation was observed after 100 continuous 2D data acquisitions at and below 150 K. Pump-probe data were used to phase only data from the 277 K flowing sample; scatter from cryogenic samples prevented accurate phasing of those datasets.

**Acknowledgements**

This work was supported by DARPA Grant HR0011-09-1-0051, and AFOSR Grant FA9550-09-1-0117 as well as funding from the Dreyfus Foundation and the Searle Foundation. Funding for J.W. and R.E.B. was provided by Grant DEFG02-07ER15846 from the Photosynthetic Systems program of the Basic Energy Sciences division of DOE.

**References**

1. Engel GS et al. (2007) Evidence for wavelike energy transfer through quantum coherence in photosynthetic systems. *Nature* 446: 782-786.
2. Lee H, Cheng YC, Fleming GR (2007) Coherence dynamics in photosynthesis: Protein protection of excitonic coherence. *Science* 316: 1462-1465.
3. Calhoun TR et al. (2009) Quantum coherence enabled determination of the energy landscape in light-harvesting complex II. *J Phys Chem B* 113: 16291-16295.
4. Beljonne D, Curutchet C, Scholes GD, Silbey RJ (2009) Beyond Förster resonance energy transfer in biological and nanoscale systems. *J Phys Chem B* 113: 6583-6599.
5. Mohseni M, Rebentrost P, Lloyd S, Aspuru-Guzik A (2008) Environment-assisted quantum walks in photosynthetic energy transfer. *J Chem Phys* 129: 9.
6. Plenio MB, Huelga SF (2008) Dephasing-assisted transport: Quantum networks and biomolecules. *New J Phys* 10: 14.
7. Rebentrost P et al. (2009) Environment-assisted quantum transport. *New J Phys* 11: 12.
8. Caruso F et al. (2009) Highly efficient energy excitation transfer in light-harvesting complexes: The fundamental role of noise-assisted transport. *J Chem Phys* 131: 15.
9. Savikhin S, Buck DR, Struve WS (1997) Pump-probe anisotropies of Fenna-Matthews-Olson protein trimers from *Chlorobium tepidum*: A diagnostic for exciton localization? *Biophys J* 73: 2090-2096.
10. Brixner T et al. (2005) Two-dimensional spectroscopy of electronic couplings in photosynthesis. *Nature* 434: 625-628.
11. Muh F et al. (2007) Alpha-helices direct excitation energy flow in the Fenna-Matthews-Olson protein. *Proc Natl Acad Sci USA* 104: 16862-16867.
12. Ishizaki A, Fleming GR (2009) Theoretical examination of quantum coherence in a photosynthetic system at physiological temperature. *Proc Natl Acad Sci USA* 106: 17255-17260.
13. van Amerongen H, Valkunas L, van Grondelle R (2000) *Photosynthetic Excitons.* (World Scientific, Singapore).
14. Blankenship RE (2002) *Molecular Mechanisms of Photosynthesis* (Blackwell Science, Oxford).

15. Fenna RE, Matthews BW (1975) Chlorophyll arrangement in a bacteriochlorophyll protein from *Chlorobium limicola*. *Nature* 258: 573-577.
16. Yang MN, Fleming GR (2003) Construction of kinetic domains in energy trapping processes and application to a photosynthetic light harvesting complex. *J Chem Phys* 119: 5614-5622.
17. Abramavicius D, Voronine DV, Mukamel S (2008) Unravelling coherent dynamics and energy dissipation in photosynthetic complexes by 2D spectroscopy. *Biophys J* 94: 3613-3619.
18. Rebentrost P, Mohseni M, Aspuru-Guzik A (2009) Role of quantum coherence and environmental fluctuations in chromophoric energy transport. *J Phys Chem B* 113: 9942-9947.
19. Collini E, Scholes GD (2009) Coherent intrachain energy migration in a conjugated polymer at room temperature. *Science* 323: 369-373.
20. Cowan ML, Ogilvie JP, Miller RJD (2004) Two-dimensional spectroscopy using diffractive optics based phased-locked photon echoes. *Chem Phys Lett* 386: 184-189.
21. Hybl JD, Ferro AA, Jonas DM (2001) Two-dimensional Fourier transform electronic spectroscopy. *J Chem Phys* 115: 6606-6622.
22. Brixner T, Mancăl T, Stiopkin IV, Fleming GR (2004) Phase-stabilized two-dimensional electronic spectroscopy. *J Chem Phys* 121: 4221-4236.
23. Cho MH (2008) Coherent two-dimensional optical spectroscopy. *Chem Rev* 108: 1331-1418.
24. Mukamel S (1995) *Principles of Nonlinear Optical Spectroscopy* (Oxford Univ Press, New York).
25. Cheng YC, Engel GS, Fleming GR (2007) Elucidation of population and coherence dynamics using cross-peaks in two-dimensional electronic spectroscopy. *Chem Phys* 341: 285-295.
26. Cheng YC, Fleming GR (2008) Coherence quantum beats in two-dimensional electronic spectroscopy. *J Phys Chem A* 112: 4254-4260.
27. Adolphs J, Renger T (2006) How proteins trigger excitation energy transfer in the FMO complex of green sulfur bacteria. *Biophys J* 91: 2778-2797.
28. Renger T, Marcus RA (2002) On the relation of protein dynamics and exciton relaxation in pigment-protein complexes: An estimation of the spectral density and a theory for the calculation of optical spectra. *J Chem Phys* 116: 9997-10019.
29. Camara-Artigas A, Blankenship RE, Allen JP (2003) The structure of the FMO protein from *Chlorobium tepidum* at 2.2 angstrom resolution. *Photosynth Res* 75: 49-55.
30. Lepetit L, Cheriaux G, Joffre M (1995) Linear techniques of phase measurement by femtosecond spectral interferometry for applications in spectroscopy. *J Opt Soc Am B* 12: 2467-2474.

**Author Contributions**